\documentclass[referee]{raa}

\usepackage{newtxtext,newtxmath}

\usepackage[T1]{fontenc}
\usepackage{ae,aecompl}

\usepackage[a4paper=true,driverfallback=dvips,pagebackref=true]{hyperref}
\hypersetup{colorlinks = true, linkcolor = green, anchorcolor = red, citecolor = blue, filecolor = red, pagecolor = red, urlcolor = red}

\usepackage{graphicx}	
\usepackage{amsmath}	

\usepackage{amssymb}	
\usepackage{natbib}

\begin{document}

\title{The possibility of detecting our solar system through astrometry}

 \volnopage{Vol.0 (20xx) No.0, 000--000}      
   \setcounter{page}{1}          

   \author{Dong-Hong, Wu
      \inst{1}
   }

   \institute{Department of Physics, Anhui Normal University, Wuhu, Anhui, 241000, China; {\it wudonghong@ahnu.edu.cn}\\
\vs\no
   {\small Received~~20xx month day; accepted~~20xx~~month day}}

\abstract{Searching for exoplanets with different methods has always been the focus of astronomers over the past few years. Among multiple planet detection techniques, astrometry stands out for its capability to accurately determine the orbital parameters of exoplanets. In this study, we examine the likelihood of extraterrestrial intelligent civilizations detecting planets in our solar system using the astrometry method. By conducting injection-recovery simulations, we investigate the detectability of the four giant planets in our solar system under different observing baselines and observational errors. Our findings indicate that extraterrestrial intelligence could detect and characterize all four giant planets, provided they are observed for a minimum of 90 years with signal-noise ratios exceeding 1. For individual planets such as Jupiter, Saturn, and Neptune, a baseline that surpasses half of their orbital periods is necessary for detection. However, Uranus requires longer observing baselines since its orbital period is roughly half of that of Neptune. If the astrometry precision is equal to or better than 10 $\mu$as, all 8,707 stars located within 30 pcs of our solar system possess the potential to detect the four giant planets within 100 years. Additionally, our prediction suggests that over 300 stars positioned within 10 pcs from our solar system could detect our Earth if they achieve an astrometry precision of 0.3 $\mu$as.
}
   \authorrunning{D.-H. Wu }            
   \titlerunning{Solar system Astrometry}  

   \maketitle


\section{Introduction}

\label{section:intro}

More than {5400} exoplanets have been detected and confirmed to date (exoplanets.nasa.gov, {July 2023}). Earth-sized habitable-zone planets turn out to orbit about one out of ten stars \citep{Petigura2013,Dressing2013}, and the search for life outside the Solar System has experienced substantial impetus. Whether a planet is habitable or not depends on how far it is from the central star and its composition \citep{Kasting1993,G_mez_Leal_2018}. Nowadays, more than 60 planets has been found to be habitable \citep{Jones2006,Lovis2006,Anglada-Escude2012,Robertson2014,Tuomi2013,Barclay2013,Borucki2013,Jenkins2015,Dittmann2017,Gilbert2023}, most of which are detected by the transit and radial velocity method. Neither the transit nor radial velocity method provides complete physical parameters of one planet, and both methods prefer to detect planets close to the central star. On the contrary, the astrometry method can provide three dimentional characterization of the orbit of one planet \citep{Perryman2014,Wu_2016} and has the advantage to detect planets far away from the host star. 

To date, only one giant planet has been detected by the Astrometry method \citep{Sahlmann2013} because of the limitation of detection precision. The detection of a habitable Earth-sized planet orbiting around a sun-like star located 10 pc away from us would require a precision of sub-$\mu$as, which is hardly achieved by the current astrometry observation such as Gaia \citep{Perryman2014}. However, it is very promising in the near future with the coming of a new era with high astrometry precision of $\mu$as \citep{Yu_2019,Ji2022,Jin2022,Tan_2022}.

Here we propose a probing question that supposing the extraterrestrial observers are using astrometry method and are also surveying the galaxy for habitable worlds, which of them could discover the planets in the solar system, even the Earth? Previous works has investigated the region in which the Earth will be observed transiting in front of the Sun \citep{Heller2016,Kaltenegger2020,Kaltenegger2021} and the frequency the Earth will be detected by other civilisations through photometric microlensing \citep{Suphapolthaworn2022}. 

In this work, we study the possibility that extraterrestrial life detect the planets in the solar system via astrometry method with {different observational precisions}. We describe how we simulate astrometric data in section \ref{sec:data_sim}. In section \ref{sec:fit}, we present how to identify planetary signals and how to fit the orbital parameters of the planets. The detection of the four giants in the solar system by nearby stars are discussed in section \ref{results}. We briefly conclude our results in section \ref{conclusion}.

\section{Simulation of Astrometric Data}\label{sec:data_sim}

Astrometry method measures the movements of the stars projected onto the celestial sky. Following the method described in previous works \citep{Black1982,Wu_2016,Yu_2019}, the projected movement of the star in the right ascension ($x$) and declination ($y$) at time $t$ can be modeled as:

\begin{equation}
    x(t)=x_0+\mu_x(t-t_0)-P_x\pi +X(t)+\sigma_x\label{eq:x}
\end{equation}
and 
\begin{equation}
    y(t)=y_0+\mu_y(t-t_0)-P_y\pi +Y(t)+\sigma_y\label{eq:y},
\end{equation}
where $x_0$ and $y_0$ are the coordinate offsets, $\mu_x$ and $\mu_y$ are the proper motions of the star, $P_x$ and $P_y$ are the parallax parameters which will be provided by the observation. $\pi$ is the annual parallax of the star. $X(t)$ and $Y(t)$ are the movements of the host star around the barycenter of the system due to the planetary companions. $\sigma_x$ and $\sigma_y$ are single-measurement astrometric errors. 

{In our fiducial simulations, we made several assumptions regarding the extraterrestrial observer's location and observational parameters. As part of our simulations, we placed the observer at a distance of 10 pcs from our Sun. The observer orbits its central star in a circular orbit with a period of 1.25 years and measures the coordinates of the Sun ($x(t)$ and $y(t)$) every 0.2 years. We also conducted simulations with data cadence of 0.1 years and found that the results changes very little.  To account for the astrometry precision, we assumed that the observer has a measurement uncertainty of 10 $\mu$as. Therefore, the individual coordinate uncertainties $\sigma_x$ and $\sigma_y$, were chosen from a Gaussian distribution with a median value of 0 and a standard deviation of 10 $\mu$as. For the coordinate offsets $x_0$ and $y_0$, we assume both to be 10 mas. Additionally, the proper motion of the Sun with respect to the observer is assumed to be 50 mas/year and -30 mas/year for the $x$ and $y$ directions, respectively. To model the parallax effect, we used the observer's orbit and defined the functions $P_x$ and $P_y$ as follows: $P_x(t)=\sin(1.6\pi t+\phi)$, $P_y(t)=\cos(1.6\pi t+\phi)$, where $\phi$ represents the orbital phase of the observer. Finally, we assumed an observing baseline of 170 years for the simulations.}

{The movement of the Sun due to the presence of the eight planets} $X(t)$ and $Y(t)$ are simulated using the REBOUND code\citep{Rein2012}. All eight planets in the Solar system are included. The orbital parameters of the planets are given by the JPL Solar System Dynamics web site\footnote{http://ssd.jp.nasa.gov/}, with respect to the mean ecliptic and equinox of J2000. In our fiducial simulation, the line of sight of the  extraterrestrial observer is assumed to be perpendicular to the mean ecliptic. {We integrate the solar system over a duration of 170 years and record the coordinates of the Sun ($X(t)$ and $Y(t)$) relative to the barycenter of the solar system every 0.2 years. }

\section{Planetary Signal identification and Orbital Parameter Fitting}\label{sec:fit}

{Assuming that we are extraterrestrial civilizations and we have measured the movement of the Sun for 170 years. Now we analyze the data to see if we have any detection. } Although we have included the gravitational interaction between planets when we simulate the astrometric data of the host star, it is ignored when we fit the orbital parameters since it has little influence on the motion of the host star \citep{Sozzetti2001,Casertano2008}. In our parameter fitting procedure, $X(t)$ and $Y(t)$ are modeled as \citep{Catanzarite2010}:
\begin{equation}
    X(t)=\sum_{i=1}^{i=N}(\cos{E_i}-e_i)A_i+\sqrt{1-e_i^2}(\sin{E_i})F_i\label{eq:X}
\end{equation}
and 
\begin{equation}
    Y(t)=\sum_{i=1}^{i=N}(\cos{E_i}-e_i)B_i+\sqrt{1-e_i^2}(\sin{E_i})G_i,\label{eq:Y}
\end{equation}
where $N$ is the number of planets orbiting around the central star, $i$ represents the $i_{\rm th}$ planet, $E_i$ is the eccentric anomaly, $e_i$ is the orbital eccentricity, $A_i$, $F_i$, $B_i$ and $G_i$ are Thiele-Innes constants, given as:

\begin{equation}
\begin{split}
A_i=\alpha_i (\cos{\Omega_i}\cos{\omega_i}-\sin{\Omega_i}\sin{\omega_i}\cos{I_i}),\\
B_i=\alpha_i(\sin{\Omega_i}\cos{\omega_i}+\cos{\Omega_i}\sin{\omega_i}\cos{I_i}),\\
F_i=\alpha_i(-\cos{\Omega_i}\sin{\omega_i}-\sin{\Omega_i}\cos{\omega_i}\cos{I_i}),\\
G_i=\alpha_i(-\sin{\Omega_i}\sin{\omega_i}+\cos{\Omega_i}\cos{\omega_i}\cos{I_i}),
\end{split}
\end{equation}

where $\alpha_i$ is the astrometric signature of the host star due to the reflex motion in the presence of the $i_{\rm th}$ planet. $\Omega$, $\omega$ and $I$ are the longitude of ascending node, arguments of pericenter and the orbital inclination of the planets, respectively. 

We search for the planetary signal and then fit the orbital parameters of the planets following the procedures as we described in \citet{Wu_2016}. Here we briefly describe the steps. 

Step 1, ignore the planetary influence on the star and use the linear least squares method to fit the five stellar parameters $x_0$, $y_0$, $\mu_x$,  $\mu_y$ and $\pi$. 

Step 2, remove the coordinate offsets, stellar proper motion and parallax from the data and search for periodical signals in the residuals using the Lomb-Scargle periodogram \citep{Black1982}. {We calculate the periodogram of the residuals in the $x$ and $y$ directions and record the most significant peak in each direction. Then we choose the peak with smaller false alarm probability (FAP). If the peak has a FAP $<10^{-4}$, we claim to have identified a certain planet signal and the corresponding orbital period is adopted as $P_1$.}

Step 3, fit $x_0$, $y_0$, $\mu_x$, $\mu_y$, $\pi$, $P_1$, $e_1$ and $t_{01}$. $t_{01}$ is the perihelion moment of the planet. This is processed via the Levenberg-Marquardt (LM) algorithm \citep{Marquardt1963} and the Markov Chain Monte Carlo (MCMC) fitting procedure. Given $P_1$, $e_1$ and $t_{01}$, the terms $\cos{E_1}-e_1$ and $\sqrt{1-e_1^2}(\sin{E_1})$ can be determined.  Then Equations \ref{eq:X} and \ref{eq:Y} are easily inverted by linear least squares to yield the Thiele-Innes constants. The motion of the Sun produced by the planet are calculated using Equation \ref{eq:X} and Equation \ref{eq:Y}. Together with the five stellar parameters, we can calculate the fitted projected motion of the star using Equation \ref{eq:x} and \ref{eq:y}. We first fit $x_0$, $y_0$, $\mu_x$, $\mu_y$, $\pi$, $P_1$, $e_1$ and $t_{01}$ using the LM method. Initial values of $x_0$, $y_0$, $\mu_x$, $\mu_y$, $\pi$ and $P_1$ are given by Step 1 and 2, while $e_1$ is randomly chosen between 0 and 1, $t_{0,1}$ is randomly chosen between 0 and $P_1$. The LM fitting process is repeated for 100 times. Then we choose the best-fit parameters with the smallest reduced $\chi^2$ as initial values of the following MCMC fitting procedure. We adopt the open-source Python package {emcee} \citep{Goodman2010,Foreman_Mackey_2013} to sample the parameter space and estimate the posterior distribution of parameters. We run emcee with 64 walkers for $20000+30000\times N$ iterations ($N$ is the number of planets identified). The initial positions of the walkers are drawn from Gaussian distributions with median values given as the best-fit parameters of the LM fitting process and standard deviations of $10^{-3}$ to accelerate the fitting process. We conduct autocorrelation analysis and find that all chains are converged in our fitting procedure.

Step 4, remove the coordinate offsets, proper motion, parallax and stellar motion due to the planet identified in Step 2 using the best-fit parameters calculated in Step 3 from the original astrometric data. Then we continue to search for periodic signals in the new residuals. If there is one, then we fit the data with a two-planet reflex motion model.

Step 5, repeat Step 2 to Step 4 until no periodic signals are identified. 

In our fitting procedure, we have a total of $5+3\times N$ parameters to fit since we have assumed Keplerian orbits for each planet, which largely reduce the parameters to be fitted and ensure the parameter precision at the same time. The semi-major axis of the planets can be obtained using the Kepler's third law giving the orbital period of the planets, while the planetary masses are calculated via $m_ia_i=m_{\odot}a_{\odot,i}$, where $m_{\odot}$ is the solar mass (assumed to be precisely determined via other methods by the extraterrestrial intelligence, like the spectrometry or astroseismology) and $a_{\odot,i}$ is the semi-major axis of the Sun when orbiting around the barycenter determined by the Sun and the $i_{\rm th}$ planet (which is obtained in Step 3). Readers are refereed to \citet{Catanzarite2010} and \citet{Wu_2016} for more detail. 

\begin{figure*}
    \centering
    \includegraphics[width=\columnwidth]{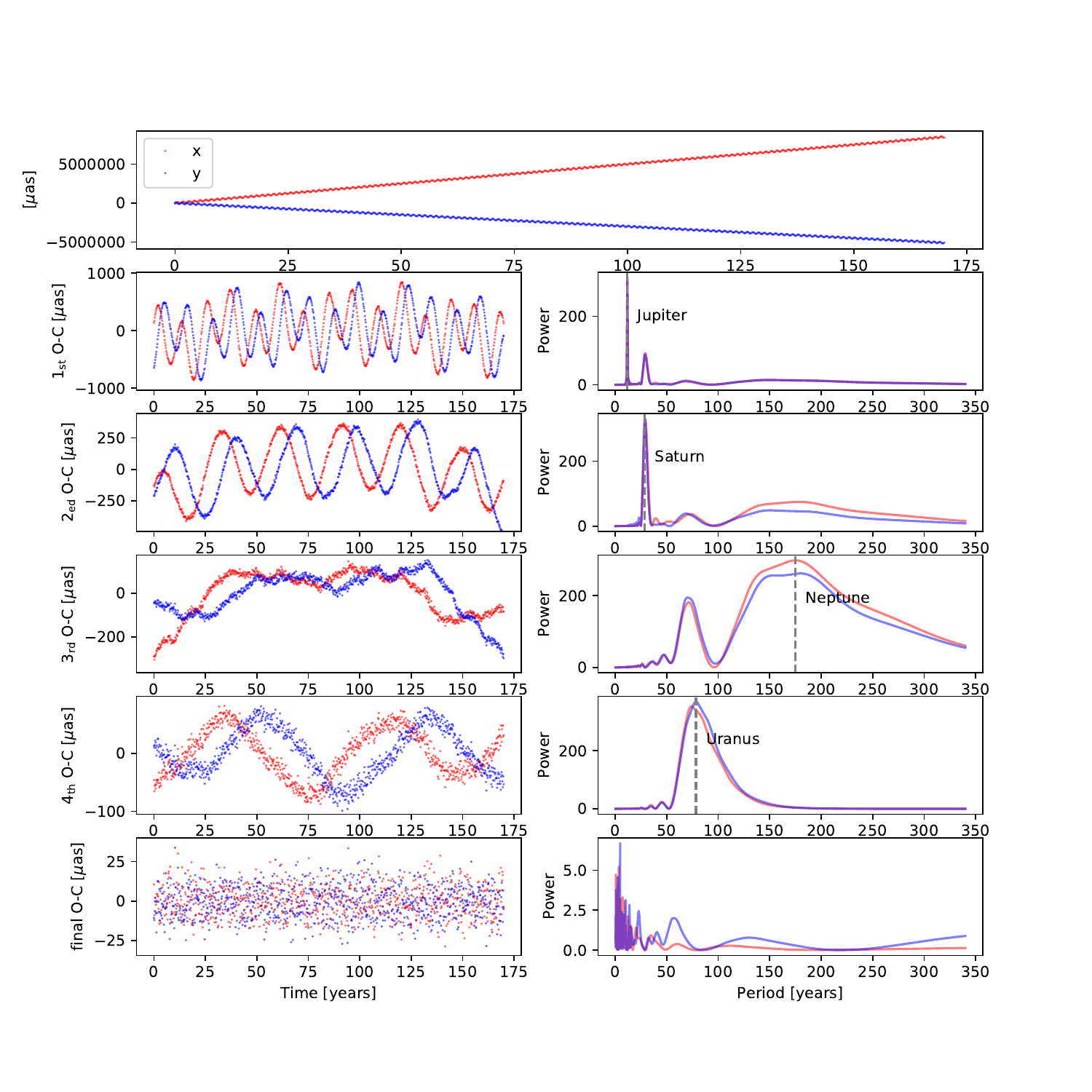}
    \caption{{Data residuals and power spectrum of the four giants after each fitting step. \textbf{Top:}The simulated astrometric data on the x (shown in red) and y (shown in blue) directions. \textbf{Left:}The data residuals after each fitting step. \textbf{Right:} The power spectrum of the data residuals shown on the left.}} 
    \label{fig:fitting_step}
\end{figure*}

{In Figure \ref{fig:fitting_step}, we present the data residuals and power spectrum for each step of the fitting process in our fiducial simulations. After Step 1 and Step 2, the power spectrum of the data residuals (labeled as $1_{\rm st}$ O-C) exhibits a prominent peak at approximately 11.9 years, indicating the successful identification of Jupiter. Then we move to Step 3 and Step 4, the updated data residuals (labeled as $2_{\rm ed}$ O-C) are also shown in Figure \ref{fig:fitting_step}, with their power spectrum peaking around 28.8 years. This peak signifies the detection of Saturn. Continuing this iterative process, we repeated the aforementioned steps, ultimately leading to the identification of Neptune and Uranus. After the detection of the four giants, no peak with FAP$<10^{-4}$ appears in the final residuals, suggesting that none of the small planets in the Solar System is detectable in our fiducial simulations.}

\section{results}\label{results}

\subsection{The characterization of the four giants in the solar system}\label{sec:detection}

The amplitude of the astrometric motion of the Sun produced by a planet with mass $m_p$ and semi-major axis $a$ observed by an observer with a distance of $d$ is:
\begin{equation}
    \alpha=3\left (\frac{m_p}{10\,m_{\oplus}} \right )\left (\frac{a}{1\, \rm AU} \right )\left ( \frac{d}{10\,\rm pc}\right)^{-1} \, \mu as. \label{snr}
\end{equation}

\begin{figure*}
    \centering
    \includegraphics[width=\columnwidth]{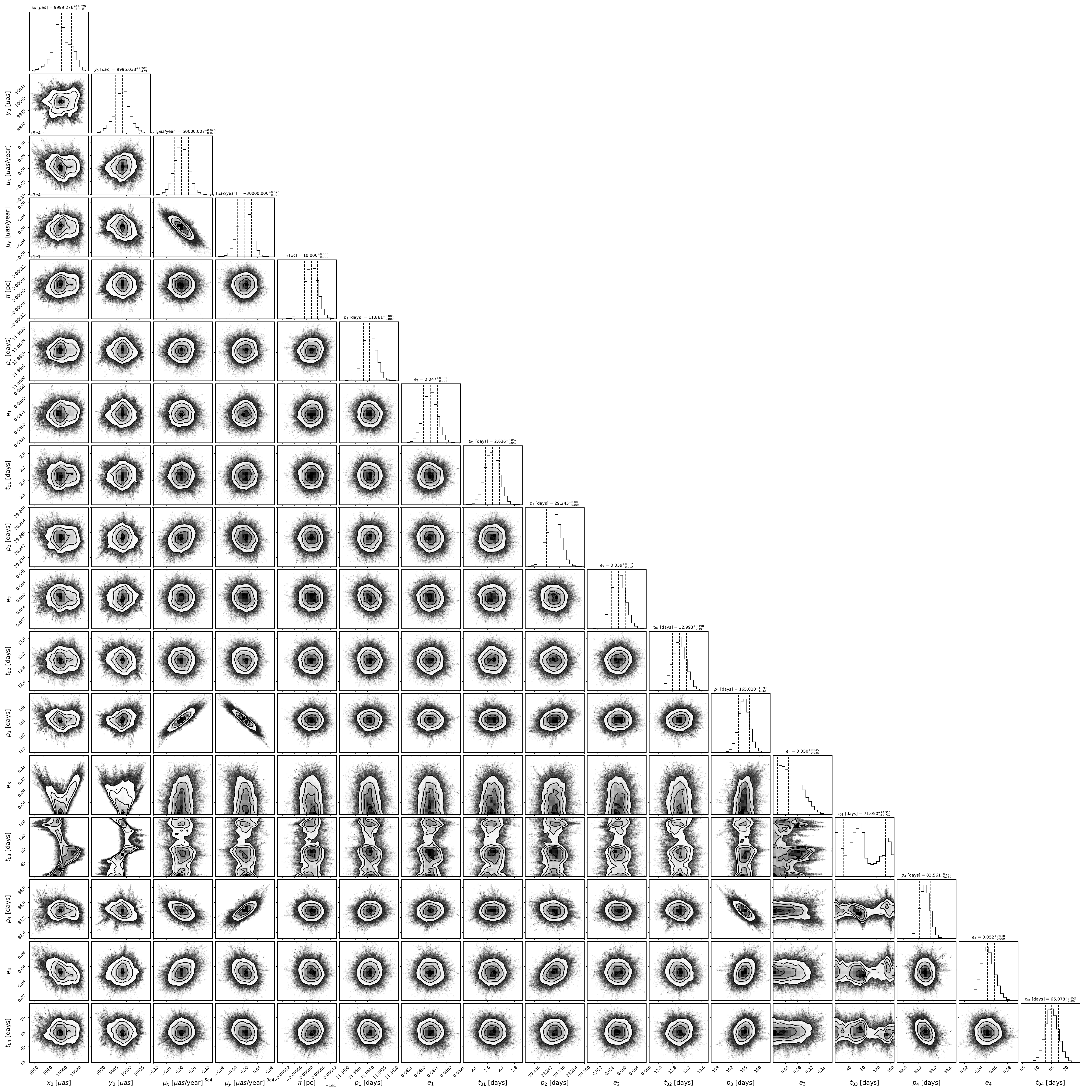}
    \caption{Posterior distribution of $x_0$, $y_0$, $\mu_x$, $\mu_y$, $\pi$, $P_1$, $e_1$, $t_{0,1}$, $P_2$, $e_2$, $t_{0,2}$, $P_3$, $e_3$, $t_{0,3}$, $P_4$, $e_4$, $t_{0,4}$. Four planets are detected by extraterrestrial intelligence located 10 pc away with an observing baseline of {170} years and observational precision of $10$ $\mu$as. } 
    \label{parameter_dis}
\end{figure*}

With an observing baseline of 170 years and observational error down to 10 $\mu$as, all the four giants are successfully detected and characterized. This is expectable since the signal-noise-ratios (SNRs, defined as $\alpha/\sigma$, where $\sigma$ is the observational error) of the four giants calculated using Equation \ref{snr} are far larger than 3, according to the detection criterion given by \citet{Wu_2016}. Other small planets in the Solar system are hardly detectable because of their small SNRs. We show the posterior distributions for all parameters that are fitted in Figure \ref{parameter_dis}. The first half iterations are thrown as burn-in. We find that the five stellar parameters ($x_0$, $y_0$, $\mu_x$, $\mu_y$ and $\pi$) and orbital parameters of the inner three giants (Jupiter, Saturn and Uranus) are all well-constrained with nearly Gaussian distributed posteriors, indicating that the parameters converge well. For the outermost planet Neptune, the orbital eccentricity ($e_3$) and perihelion moment ($t_{03}$) are not well-constrained since the planet only finishes one complete orbit during 170 years. The planetary mass and semi-major axis of the planets can be easily calculated using the best-fit parameters as we have mentioned in section \ref{sec:fit}. All the four giants are well characterized with relative fitting errors smaller than $1\%$ for both the orbital period and planet mass. The relative fitting error of parameter $\theta$ is given as $\epsilon_{\theta}=|\theta_{\rm fit}-\theta_{\rm true}|/\theta_{\rm true}$. $\theta_{\rm fit}$ is calculated as the median value of the posterior distribution of $\theta$. 

\subsection{The influence of observing baseline and observational error}

We also investigated the detection of the four giants with different observational errors and observing baselines. In our fiducial simulations described in Section \ref{sec:data_sim}, we fix the observing baseline to be 170 years and the observational error to be 10 $\mu$as. Now we gradually decrease the observing baseline from 170 years to 10 years, with a step of 20 years. {To account for the detection of our Solar system by missions like Gaia \citep{Perryman2014} and CHES \citep{Ji2022}, we further extend the observing baseline down to 4 years. }We also considered different observational errors: 1 $\mu$as, 3 $\mu$as, 10 $\mu$as, 30 $\mu$as, 100 $\mu$as, 300 $\mu$as, 1000 $\mu$as, 3000 $\mu$as and 10000 $\mu$as. Other assumptions such as the distance of the observer and the sampling cadence remain the same. In the new simulations, we assume that the coordinates offsets, proper motion and parallax of our Sun are already well determined and carefully removed from the astrometric data by extraterrestrial intelligence before the fitting process starts. This assumption will largely reduce the computational time (We have conducted a small group of simulations including the coordinate offsets, proper motion and parallax, and we find that they have little influence on the characterization of the planets). Then we start the fitting procedure as described in Section \ref{sec:fit} but now we skip step 1. 

\begin{figure*}
    \centering
    \includegraphics[width=\columnwidth]{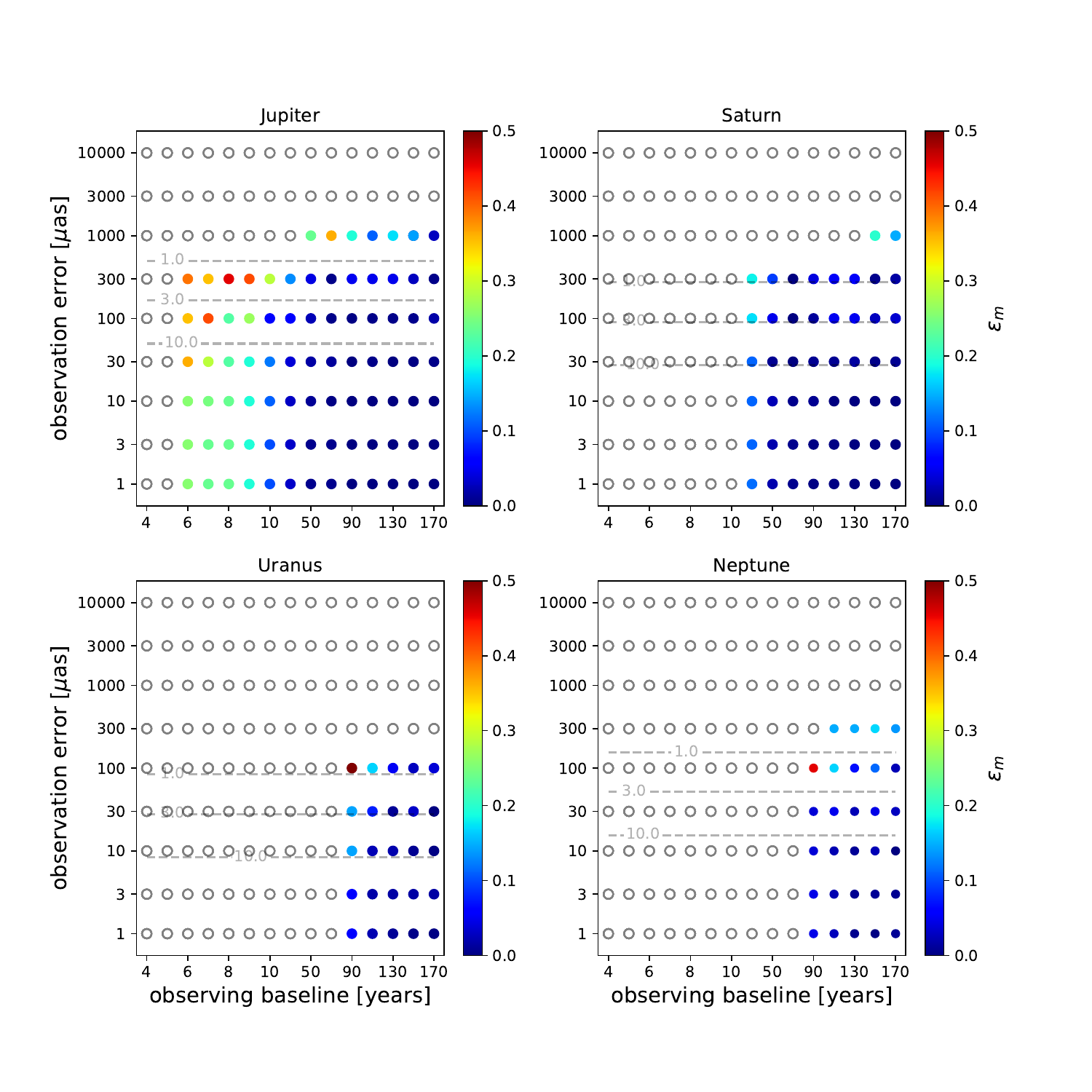}
    \caption{{The relative fitting errors of the planet mass $\epsilon_m$ as a function of observing baseline and observational error. Different colors represent different $\epsilon_m$. Gray circles represent none detection of planet signals or the identified planet has large fitting errors of orbital period (i.e. $\epsilon_P>0.1$). The gray dashed lines represent the contours of different SNRs.  } }
    \label{planet_err}
\end{figure*}

We show the relative fitting errors of planet mass as a function of the observing baseline and observational error for each of the four giants in Figure \ref{planet_err}. Only planets with relative fitting errors of orbital period smaller than 0.1 ($\epsilon_P<0.1$) are shown. For planets with $\epsilon_P<0.1$, their planetary mass are mostly well-fitted with $\epsilon_m<0.1$. Only a few exceptions with large observational errors or short observing baselines have large fitting errors of planet mass. There are several cases that planets are detected with $\epsilon_P>0.1$, however, their planet mass are generally poorly fitted with $\epsilon_m>1$. Therefore, we claim a planet is well characterized if it has $\epsilon_P<0.1$.

As we have pointed out in \citet{Wu_2016}, the detection of a planet using astrometry method relys on the SNR of the planet and the observing baseline. Here we show the contours of the SNRs in Figure \ref{planet_err}. We find that all the four giants can be successfully detected and well-characterized as long as their SNRs $>1$ and the observing baseline exceeds 90 years. In general, planets with SNRs $>1$ and observing baseline longer than about half an orbital period could be detected. However, this is not the case for Uranus. Because the fitting of the orbital period of Uranus is largely influenced by that of Neptune, whose orbital period is about two times that of Uranus. Not until the orbital period of Neptune is successfully identified will Uranus be well-characterized. There are exceptions that planets are detected with SNR $<1$. However, these detections generally have large fitting errors of planet mass.

{If astrometric missions conducted by extraterrestrial civilizations are similar to Gaia or CHES, with typical observing baselines ranging from 5 to 10 years, the detectability of the giant planets in our simulations would be limited. Specifically, only Jupiter would be detectable  under these circumstances. However, if Gaia were able to achieve an observational error down to 100 $\mu$as, it would be possible to characterize Jupiter with an accuracy of $\epsilon_m<0.5$. Alternatively, if CHES, with an observational error down to 1 $\mu$as and and an observational time of approximately 6 years, conducted the mission, Jupiter could be characterized with an accuracy of $\epsilon_m \sim 0.3$.}

\subsection{Which stars could detect the four giants in the solar system}

{We move forward to estimate how many neighbouring stars in the Galaxy could detect the four giants in our solar system. We identify 8707 stars from the Gaia Catalog of Nearby Stars (GCNS)\citep{Gaia_Collaboration2021} that lie within 30 pc to the Solar system. We calculate the SNR of each giant planet observed by each star with different assumed observational errors. We find that all 8707 stars have the possibility to detect and well-characterize the four giants if they could achieve an astrometric error down to 10 $\mu$as and observe the solar system for enough long time (such as 90 years). If the observational error is as large as 100 $\mu$as, only 183 neighbouring stars could detect all the four giants, but all of them could detect the Jupiter within 10 years. We also estimate the number of neighbouring stars that could detect our Earth. About 310 neighbouring stars located within 10 pc from our Sun have the potential to detect the Earth if the observational error is as small as $0.3$ $\mu$as. With a larger observational error, such as 1 $\mu$as, only 8 stars located within 3 pc from the Sun could possibly detect the Earth. }

\section{Conclusion}\label{conclusion}

In this paper, we study the possibility that extraterrestrial intelligence detect the planets in our solar system. {We find that all the four giants in our solar system could be detected and well-characterized as long as they are observed for at least 90 years with SNR $>1$. For all 8707 stars lying within 30 pc to the solar system, all of them have the potential to detect the four giants within 100 years if they could achieve an observational precision down to 10 $\mu$as. If the astrometry method can achieve sub $\mu$as precision like 0.3 $\mu$as, then even our Earth will be detectable by extraterrestrial intelligence.}

In each of our simulations, we assume a constant observational error during the long observing baseline for simplicity. A more reasonable assumption maybe a decreasing observational error with the increase of observing baseline. {Besides, the sampling cadence is fixed to be 0.2 years in our simulations. This is hardly achieved in real observations. Finally, our simulations is truncated at 170 years since longer observing baseline requires longer computational time to fit the orbital parameters of the planets. However, we expect that longer observing baseline would allow the detection of plants with larger observational errors. These should be further considered in future works.}

{This study primarily addresses the likelihood of extraterrestrial civilizations in the vicinity of our solar system detecting our own system. However, it is important to note that the existence of extraterrestrial life remains uncertain, and if they do exist in planetary systems similar to ours, their presence could be incredibly rare. \citet{Cumming_2008} demonstrated that the occurrence rate of cold Jupiters around stars similar to the Sun is only $10\%$. Consequently, the chances of our solar system being in proximity to a significant population of extraterrestrial civilizations are currently considered to be very small. 
 }
\section{Acknowledgements}
\label{section:acknowledgements}

This work is supported by the National Natural Science Foundation of China (NSFC) (grant No. 12103003), and the Doctoral research start-up funding of Anhui Normal University. 

\bibliographystyle{raa}
\bibliography{cite}

\begin{thebibliography}{35}
\providecommand\natexlab[1]{#1}
\providecommand\JournalTitle[1]{#1}

\bibitem[{Anglada-Escud{\'e}} {et~al.}(2012)]{Anglada-Escude2012}
{Anglada-Escud{\'e}}, G., {Arriagada}, P., {Vogt}, S.~S., {et~al.} 2012, \apjl,
  751, L16

\bibitem[{Barclay} {et~al.}(2013)]{Barclay2013}
{Barclay}, T., {Burke}, C.~J., {Howell}, S.~B., {et~al.} 2013, \apj, 768, 101

\bibitem[{Black} \& {Scargle}(1982)]{Black1982}
{Black}, D.~C., \& {Scargle}, J.~D. 1982, \apj, 263, 854

\bibitem[{Borucki} {et~al.}(2013)]{Borucki2013}
{Borucki}, W.~J., {Agol}, E., {Fressin}, F., {et~al.} 2013, Science, 340, 587

\bibitem[{Casertano} {et~al.}(2008)]{Casertano2008}
{Casertano}, S., {Lattanzi}, M.~G., {Sozzetti}, A., {et~al.} 2008, \aap, 482,
  699

\bibitem[{Catanzarite}(2010)]{Catanzarite2010}
{Catanzarite}, J.~H. 2010, arXiv e-prints, arXiv:1008.3416

\bibitem[Cumming {et~al.}(2008)]{Cumming_2008}
Cumming, A., Butler, R.~P., Marcy, G.~W., {et~al.} 2008, Publications of the
  Astronomical Society of the Pacific, 120, 531

\bibitem[{Dittmann} {et~al.}(2017)]{Dittmann2017}
{Dittmann}, J.~A., {Irwin}, J.~M., {Charbonneau}, D., {et~al.} 2017, \nat, 544,
  333

\bibitem[{Dressing} \& {Charbonneau}(2013)]{Dressing2013}
{Dressing}, C.~D., \& {Charbonneau}, D. 2013, \apj, 767, 95

\bibitem[Foreman-Mackey {et~al.}(2013)]{Foreman_Mackey_2013}
Foreman-Mackey, D., Hogg, D.~W., Lang, D., \& Goodman, J. 2013, Publications of
  the Astronomical Society of the Pacific, 125, 306

\bibitem[{Gaia Collaboration} {et~al.}(2021)]{Gaia_Collaboration2021}
{Gaia Collaboration}, {Smart}, R.~L., {Sarro}, L.~M., {et~al.} 2021, \aap, 649,
  A6

\bibitem[{Gilbert} {et~al.}(2023)]{Gilbert2023}
{Gilbert}, E.~A., {Vanderburg}, A., {Rodriguez}, J.~E., {et~al.} 2023, \apjl,
  944, L35

\bibitem[G{\'{o}}mez-Leal {et~al.}(2018)]{G_mez_Leal_2018}
G{\'{o}}mez-Leal, I., Kaltenegger, L., Lucarini, V., \& Lunkeit, F. 2018, The
  Astrophysical Journal, 869, 129

\bibitem[{Goodman} \& {Weare}(2010)]{Goodman2010}
{Goodman}, J., \& {Weare}, J. 2010, Communications in Applied Mathematics and
  Computational Science, 5, 65

\bibitem[Heller \& Pudritz(2016)]{Heller2016}
Heller, R., \& Pudritz, R.~E. 2016, Astrobiology, 16, 259, pMID: 26967201

\bibitem[{Jenkins} {et~al.}(2015)]{Jenkins2015}
{Jenkins}, J.~M., {Twicken}, J.~D., {Batalha}, N.~M., {et~al.} 2015, \aj, 150,
  56

\bibitem[{Ji} {et~al.}(2022)]{Ji2022}
{Ji}, J.-H., {Li}, H.-T., {Zhang}, J.-B., {et~al.} 2022, Research in Astronomy
  and Astrophysics, 22, 072003

\bibitem[{Jin} {et~al.}(2022)]{Jin2022}
{Jin}, S., {Ding}, X., {Wang}, S., {Dong}, Y., \& {Ji}, J. 2022, \mnras, 509,
  4608

\bibitem[{Jones} {et~al.}(2006)]{Jones2006}
{Jones}, B.~W., {Sleep}, P.~N., \& {Underwood}, D.~R. 2006, \apj, 649, 1010

\bibitem[{Kaltenegger} \& {Faherty}(2021)]{Kaltenegger2021}
{Kaltenegger}, L., \& {Faherty}, J.~K. 2021, \nat, 594, 505

\bibitem[{Kaltenegger} \& {Pepper}(2020)]{Kaltenegger2020}
{Kaltenegger}, L., \& {Pepper}, J. 2020, \mnras, 499, L111

\bibitem[{Kasting} {et~al.}(1993)]{Kasting1993}
{Kasting}, J.~F., {Whitmire}, D.~P., \& {Reynolds}, R.~T. 1993, \icarus, 101,
  108

\bibitem[{Lovis} {et~al.}(2006)]{Lovis2006}
{Lovis}, C., {Mayor}, M., {Pepe}, F., {et~al.} 2006, \nat, 441, 305

\bibitem[Marquardt(1963)]{Marquardt1963}
Marquardt, D.~W. 1963, Journal of the Society for Industrial and Applied
  Mathematics, 11, 431

\bibitem[{Perryman} {et~al.}(2014)]{Perryman2014}
{Perryman}, M., {Hartman}, J., {Bakos}, G.~{\'A}., \& {Lindegren}, L. 2014,
  \apj, 797, 14

\bibitem[{Petigura} {et~al.}(2013)]{Petigura2013}
{Petigura}, E.~A., {Howard}, A.~W., \& {Marcy}, G.~W. 2013, Proceedings of the
  National Academy of Science, 110, 19273

\bibitem[{Rein} \& {Liu}(2012)]{Rein2012}
{Rein}, H., \& {Liu}, S.~F. 2012, \aap, 537, A128

\bibitem[{Robertson} {et~al.}(2014)]{Robertson2014}
{Robertson}, P., {Mahadevan}, S., {Endl}, M., \& {Roy}, A. 2014, Science, 345,
  440

\bibitem[{Sahlmann} {et~al.}(2013)]{Sahlmann2013}
{Sahlmann}, J., {Lazorenko}, P.~F., {S{\'e}gransan}, D., {et~al.} 2013, \aap,
  556, A133

\bibitem[{Sozzetti} {et~al.}(2001)]{Sozzetti2001}
{Sozzetti}, A., {Casertano}, S., {Lattanzi}, M.~G., \& {Spagna}, A. 2001, \aap,
  373, L21

\bibitem[{Suphapolthaworn} {et~al.}(2022)]{Suphapolthaworn2022}
{Suphapolthaworn}, S., {Awiphan}, S., {Chatchadanoraset}, T., {et~al.} 2022,
  arXiv:2206.09820

\bibitem[Tan {et~al.}(2022)]{Tan_2022}
Tan, D.-J., Liu, J.-C., Zhu, Z., \& Liu, N. 2022, Research in Astronomy and
  Astrophysics, 22, 025008

\bibitem[{Tuomi} {et~al.}(2013)]{Tuomi2013}
{Tuomi}, M., {Anglada-Escud{\'e}}, G., {Gerlach}, E., {et~al.} 2013, \aap, 549,
  A48

\bibitem[Wu {et~al.}(2016)]{Wu_2016}
Wu, D.-H., Liu, H.-G., Yu, Z.-Y., Zhang, H., \& Zhou, J.-L. 2016, The
  Astrophysical Journal, 825, 76

\bibitem[Yu {et~al.}(2019)]{Yu_2019}
Yu, Z.-Y., Liu, H.-G., Zhou, J.-L., {et~al.} 2019, Research in Astronomy and
  Astrophysics, 19, 004

\end{thebibliography}







\label{lastpage}
\end{document}